\documentclass[preprint]{elsarticle}

\usepackage{euscript,amssymb,amsmath}
\usepackage{amsfonts,latexsym}
\usepackage{amsfonts,amsmath,amssymb,amsfonts,latexsym}
\usepackage[utf8]{inputenc}

\usepackage{graphicx}
\usepackage{epsfig}
\usepackage{color}

\newcommand{\be}[1]{\begin{equation}\label{#1}}
\newcommand{\ee}{\end{equation}}
\newcommand{\ba}[1]{\begin{eqnarray}\label{#1}}
\newcommand{\ea}{\end{eqnarray}}
\newcommand{\rf}[1]{(\ref{#1})}
\newcommand{\nn}{\nonumber}

\newcommand{\sfrac}[2]{\dfrac{\,#1\,}{\,#2\,}}

\newcommand{\der}[2]{\sfrac{\partial #1}{\partial #2}}

\newcommand{\dder}[2]{\sfrac{\partial^{\,2} #1}{\partial #2^2}}

\def\H{{\cal H}}      
\def\K{{\cal K}}      

\newcommand{\al}{\alpha}
\newcommand{\bt}{\beta}

\newcommand{\gm}{\gamma}

\newcommand{\dl}[1]{\partial_{#1}}
\newcommand{\mf}[1]{\mathbf{#1}}

\begin{document}
	
\begin{frontmatter}
		
\title{Effect of peculiar velocities of inhomogeneities\\ on the shape of gravitational potential\\ in spatially curved universe}
		
\author[a]{\,Ezgi Canay\,}
\ead{ezgicanay@itu.edu.tr}

\author[b]{\,Maxim Eingorn\,}
\ead{maxim.eingorn@gmail.com}

\author[b]{\,Andrew McLaughlin II\,}
\ead{amclau12@eagles.nccu.edu}

\author[a]{\,A.~Sava\c{s}~Arapo\u{g}lu\,}
\ead{arapoglu@itu.edu.tr}
		
\author[c,d]{\,Alexander Zhuk\,}
\ead{ai.zhuk2@gmail.com}

\address[a]{Department of Physics, Istanbul Technical University,\\ 34469 Maslak, Istanbul, Turkey}

\address[b]{Department of Mathematics and Physics, North Carolina Central University,\\ 1801 Fayetteville St., Durham, North Carolina 27707, U.S.A.}

\address[c]{Astronomical Observatory, Odessa National University,\\ Dvoryanskaya st. 2, Odessa 65082, Ukraine}
		
\address[d]{Center for Advanced Systems Understanding (CASUS), Untermarkt 20, 02826 G{\"o}rlitz, Germany}

\begin{abstract} We investigate the effect of  peculiar velocities of inhomogeneities and the spatial curvature of the universe  on the shape of the gravitating potential. To this end, we consider scalar perturbations of the FLRW metric. The gravitational potential satisfies a Helmholtz-type equation which follows from the system of linearized Einstein equations.	
We obtain analytical solutions of  this equation in the cases of open and closed universes, filled with cold dark matter in presence of the cosmological constant. We demonstrate that, first, peculiar velocities significantly affect the screening length of the gravitational interaction and, second,  the form of the gravitational potential depends on the sign of the spatial curvature.
\end{abstract}

		
\end{frontmatter}
	
\section{Introduction}
	
\setcounter{equation}{0}

The density parameter associated with the spatial curvature of the universe is of great importance in cosmological studies. For the homogeneous and isotropic model, it reads $\Omega_{\mathcal{K}}\equiv -\mathcal{K} c^2/(a_0^2H_0^2)$, where $c$ is the speed of light, $a_0$ and $H_0$ are the scale factor and Hubble parameter at the present time. The value of $\mathcal{K} = -1, 0, +1$ for open, flat and closed topologies, respectively. In the standard $\Lambda$CDM model, $\mathcal{K}=0$, i.e. the universe is spatially flat. Recent cosmological observations, e.g., combined Planck 2018 cosmic microwave background and baryon acoustic oscillation measurements, strongly favor this value \cite{Planck}. However,  the combination of the Planck temperature and polarization power spectra suggests $\Omega_{\mathcal{K}}=-0.044^{+0.018}_{-0.015}$,  favoring a mildly closed universe.  Hence, it can be readily concluded that observations do not yet give an unambiguous answer regarding the value of the spatial curvature parameter. This problem is the subject of intensive modern research (see, e.g., recent papers \cite{K1,K2,K3,K4,K5,K6,K7,K8}) and the importance of these studies lies in the fact that the curvature parameter affects, e.g.,  the global dynamics of the universe, the lensing \cite{lensing}, and the form of metric perturbations, in particular, the shape of the gravitational potential.     

In our previous papers \cite{Burgazli1,Emrah}, we have shown that the gravitational potential has different forms for the open, flat and closed universe cases. However, both these papers have an essential drawback: peculiar velocities of inhomogeneities such as galaxies and groups of galaxies are disregarded. Nevertheless, it has been shown in \cite{EE} that peculiar velocities have a considerable impact on the screening cutoff scale of the gravitational interaction for the flat universe, filled with cold dark matter in the presence of the cosmological constant $\Lambda$. In the case of perfect fluids with constant equation of state parameter $\omega$ (in particular, for $\omega =\pm1/3$), such an effect has been studied separately in \cite{Burgazli2}. 

The main goal of the present article is to investigate in which way the spatial curvature of the universe and peculiar velocities of inhomogeneities affect the behaviour of gravitational interactions. In section 2, we begin by describing the background model. Then, in section 3, we consider scalar perturbations of the background and derive the equation for the gravitational potential. We demonstrate that this equation is of Helmholtz type and hence admits a characteristic screening length. In section 4, we obtain solutions of this equation in the cases of spatially open and closed universes. In concluding section 5, we discuss the obtained results.

\section{Background  model}

\setcounter{equation}{0}

The background geometry is described by the Friedmann-Lema$\mathrm{\hat{\i}}$tre-Robertson-Walker (FLRW) metric
\ba{2.1}
ds^2 = a^2(\eta) \big[ d\eta^2 - \gm_{\al \bt} \, dx^\al dx^\bt \,\big] = a^2(\eta) \big[ d\eta^2 - d\chi^2 - \Sigma^2(\chi)  \,d\Omega^2 \big]\, , \ea
where $a(\eta)$ is the scale factor while $\eta$ is the conformal time ($\eta$ is related to the synchronous time $t$ via $d\eta = cdt/ a$). The function $\Sigma(\chi)$ reads
\be{2.2}
\Sigma(\chi) =
\begin{cases}
	\sin\!\chi, &\quad \chi \in [0,\pi] \;\text{for}\; \K=+1 \\
	\chi, &\quad \chi \in [0,+\infty) \;\text{for}\; \K=0 \\
	\sinh\!\chi, &\quad \chi \in [0,+\infty) \;\text{for}\; \K=-1 \\
\end{cases}
\ee 
where $\K=-1,0,+1$ for open, flat and closed universes, respectively.

We consider the universe filled with 
nonrelativistic pressureless matter (e.g., cold dark matter) with the average energy density \mbox{$\bar\varepsilon = \bar{\rho}c^2/a^3$}, $\bar\rho$ being the average comoving mass density. Then, the Friedmann equation reads
\be{2.3}
\sfrac{3(\H^2 + \K)}{a^2} = \kappa \bar{\varepsilon} + \Lambda\, ,
\ee
in which $\H\equiv (da/d\eta)/a =(a/c)H$ and $H\equiv (da/dt)/a$ is the Hubble parameter. We also define $\kappa \equiv 8\pi G_{\!N}/c^4$, where $G_{\!N}$ denotes the gravitational constant.

For the given background model, the dimensionless cosmological parameters are
\be{2.4}
\Omega_{\rm M} \equiv \sfrac{\kappa \bar{\rho} c^4}{3 H_0^2 a_0^3} \;, \quad \Omega_\Lambda \equiv \sfrac{\Lambda c^2}{3 H_0^2} \;, \quad
\Omega_\K \equiv -\sfrac{\K c^2}{a_0^2 H_0^2}\, .
\ee
In what follows, we will use the results of \cite{Planck} for the values of these cosmological parameters. 
With the help of these expressions, the Friedmann equation \rf{2.3}  can be rearranged as 
\be{2.5} H=H_0\sqrt{\Omega_M\left(\frac{a_0}{a}\right)^3+\left(1-\Omega_M-\Omega_{\Lambda}\right)\left(\frac{a_0}{a}\right)^2+\Omega_{\Lambda}} \, .
\ee


\section{Scalar perturbations}

We consider the matter component (e.g., galaxies) in the form of discrete point-like masses. Their comoving mass density reads
\be{2.6}
\rho = \sum_i \rho_i = \sfrac{1}{\sqrt{\gm}} \sum_i m_i \, 
\delta(\mf{r}-\mf{r}_i) \, ,
\ee
where $\gm$ denotes the determinant of $\gm_{\alpha\beta}$.  These discrete inhomogeneities perturb the background metric \rf{2.1} to yield 
\be{2.7} ds^2 = a^2 \big[ (1+2\Phi) d\eta^2  - (1-2\Phi) \gm_{\al\bt} \, dx^\al dx^\bt\,\big]\, , \ee
provided that we confine ourselves to scalar perturbations only. It is well known that the first-order scalar perturbation $\Phi\left(\eta,\mathbf{r}\right)$ is the gravitational potential at the point with the radius-vector $\mathbf{r}$, produced by all inhomogeneities in the system. The linearized Einstein equations of the relativistic perturbation theory read \cite{Mukhanov}
\be{2.8}\triangle\Phi-3\mathcal{H}\left(\Phi^\prime+\mathcal{H}\Phi\right)+3\mathcal{K}\Phi=\frac{1}{2}\kappa a^2\delta\varepsilon
\, ,\ee
\be{2.9}
\Phi'+\mathcal{H}\Phi=-\frac{1}{2}\kappa a^2\overline{\varepsilon}v\, ,\ee
\be{2.10}
\Phi''+3\mathcal{H}\Phi'+\left(2\mathcal{H}'+\mathcal{H}^2-\mathcal{K}\right)\Phi=0\, ,
\ee
where the prime denotes the $\eta$-derivative while $\Delta = \left({1}/{\sqrt{\gamma}}\right) \dl{\alpha} \big(\sqrt{\gamma}\,\gamma^{\alpha\beta}\dl{\beta}\big)$ is the Laplace operator. On the right-hand sides, $\delta\varepsilon$ represents the energy density fluctuation while $v$ stands for the velocity potential.

Following \cite{ansatz}, we use the ansatz 
\be{2.11}
\Phi\left(\eta,\mathbf{r}\right)=\frac{D_1(\eta)}{a(\eta)}\phi(\mathbf{r}) \, .
\ee
Then, substituting
\be{2.12}
\Phi'+\mathcal{H}\Phi=\frac{D'_1}{D_1}\Phi \, 
\ee
into \rf{2.8}, we get 
\be{2.13}
\triangle\Phi-3\left(\mathcal{H}\frac{D'_1}{D_1}-\mathcal{K}\right)\Phi=\frac{1}{2}\kappa a^2\delta\varepsilon \, .
\ee
Using the relation $\delta\varepsilon=c^2\delta\rho/a^3 +3\overline\rho c^2\Phi /a^3$ (see, e.g., \cite{EZflow,EZremarks,Ein1}), we may rearrange \rf{2.13} to obtain
\be{2.14}
\triangle\Phi-3\left(\mathcal{H}\frac{D'_1}{D_1}+\frac{\kappa\overline\rho c^2}{2a}-\mathcal{K}\right)\Phi= \frac{\kappa c^2}{2a}\delta\rho \, ,
\ee
or
\be{2.15}
\triangle\Phi-\frac{a^2}{\lambda^2_{\rm eff}}\Phi= \frac{\kappa c^2}{2a}\delta\rho \, .
\ee
For this Helmholtz-type equation, we introduce the effective screening length:
\be{2.16}
\frac{1}{\lambda^2_{\rm eff}}\equiv\frac{1}{\lambda^2} +\frac{1}{a^2l^2}\, .
\ee
Here,
\be{2.17}
\frac{1}{\lambda^2}\equiv \frac{3\kappa \overline\rho c^2}{2a^3}
\ee
and 
\be{2.18}
\frac{1}{l^2}\equiv  3\mathcal{H}\frac{D'_1}{D_1}-3\mathcal{K}
=3\mathcal{H}^2 \frac{d\ln D_1}{d\ln a}  -3\mathcal{K} \, .
\ee
The second term on the right-hand side of Eq.~\rf{2.16} incorporates the effects of both the curvature and peculiar velocities on the screening length.

Now, substituting the ansatz \rf{2.11} into Eq.~\rf{2.10}, we obtain
\be{2.19}
D^{\prime\prime}_1+\mathcal{H}D'_1+\left(\mathcal{H}'-\mathcal{H}^2-\mathcal{K}\right)D_1=0 \,,
\ee
with two independent solutions \cite{0112551}:
\be{2.20}
D^{(+)}_1\propto H\int\frac{da}{(aH)^3}\propto \frac{\mathcal{H}}{a}\int\frac{da}{\mathcal{H}^3} \, 
\ee
and
\be{2.21}
D^{(-)}_1\propto \frac{\mathcal{H}}{a} \, .
\ee
The second solution corresponds to the decaying mode and hence will not be of interest for us. 

Substituting $D^{(+)}_1$ into \rf{2.16}, we get
\ba{2.22} \frac{1}{\lambda_{\mathrm{eff}}^2}
&=&\frac{3\kappa \overline\rho c^2}{2a^3}+\frac{3\mathcal{H}}{a^2}\left[\frac{a}{\mathcal{H}^2}\left(\int\frac{da}{\mathcal{H}^3}\right)^{-1}+\left(a\frac{d\mathcal{H}}{da}-\mathcal{H}-\frac{\mathcal{K}}{\mathcal{H}}\right)\right] \nn\\
&=&
\frac{3}{Ha^2c^2}\left(\int \frac{da}{a^3H^3}\right)^{-1}
\, ,
\ea
where we use Eq. \rf{2.3} to simplify the right-hand side.


\section{Gravitational potential}

\begin{figure*}[!ht]
	\centering
	
	\begin{tabular}{@{}c@{}}
		\includegraphics[width=\linewidth]{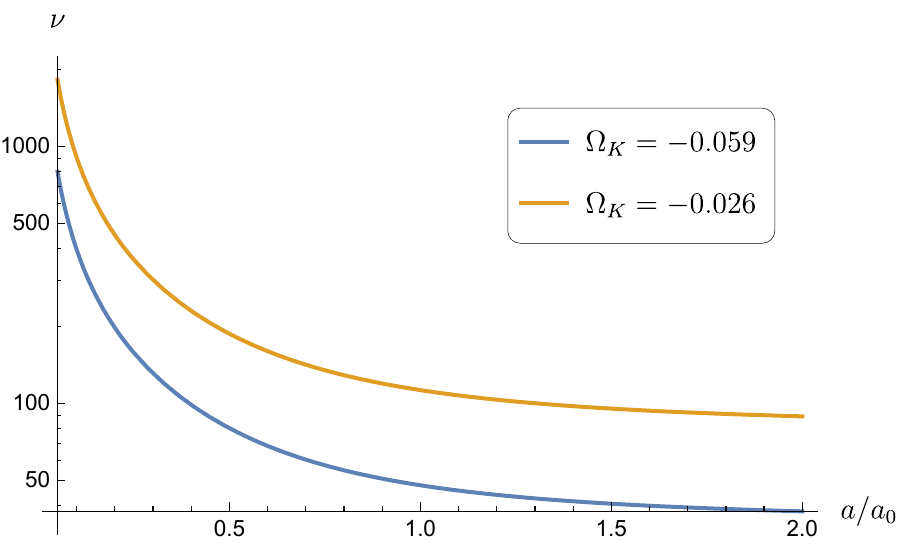} \\
	\end{tabular}
	
	\vspace{-2mm}
	
	\caption{Behaviour of the parameter $\nu$ as a function of the scale factor. The top (orange) and bottom (blue) lines correspond, respectively, to the upper  and lower limits according to the Planck 2018 measurements. The parameter $\nu$ is significantly greater than 1 at all the evolution stages of interest for the considered universe model.
	}
	\label{fig.1}
\end{figure*}

Now, returning to Eq.~\rf{2.15}, we find it convenient to introduce the new function
\be{2.23}  
\phi=c^2a\Phi-\frac{4\pi G_N\overline\rho}{\nu}\, ,
\ee
which satisfies the equation
\be{2.24}
\triangle\phi-\nu\phi= 4\pi G_N\rho\, ,
\ee
where $\rho$ is given by Eq.~\rf{2.6} and
\be{2.25}
\nu\equiv \frac{a^2}{\lambda^2_{\rm eff}}>0\, .
\ee
The potential $\phi$ satisfies the superposition principle and a total solution of \rf{2.24} can be found by summing the individual potentials $\phi_i$, sourced by the corresponding masses $m_i$. For the $i$-th mass located at the origin of coordinates, the function $\phi_i$ satisfies 
\be{2.26}
\sfrac{1}{\Sigma^2(\chi)} \der{}{\chi} \bigg( \Sigma^2(\chi) \,\der{\phi_i}{\chi} \bigg) - \nu \phi_i = 0\, .
\ee
Following a redefinition $U(\eta,\chi) \equiv \Sigma(\chi)\,\phi_i(\eta,\chi)$, this equation is considerably simplified to yield
\be{2.27}
\dder{U}{\chi} - \mu U = 0\, ,
\ee
where
\be{2.28}
\mu \equiv \sfrac{1}{\Sigma(\chi)} \dder{\Sigma(\chi)}{\chi} + \nu =
\begin{cases}
	\nu-1 & \;\text{for}\; \K=+1 \\
	\nu+1 & \;\text{for}\; \K=-1 \\
\end{cases}
\ee
The form of the solution of Eq.~\rf{2.27} depends on the sign of the parameter $\mu$. This parameter is positive in the case of an open universe.
As regards the closed universe case, during all the stages of the evolution relevant to our investigation (i.e. when the dominant contribution to the energy density comes from cold dark matter plus $\Lambda$), $\nu$ is much bigger than unity. Figure \ref{fig.1} clearly confirms this statement. Here we have used $\Omega_{\mathcal{K}}=-0.044^{+0.018}_{-0.015}$. The top (orange) and bottom (blue) lines correspond to the upper and lower limits, respectively.

\vspace{1cm}

\

Therefore, for the closed universe, we will consider only the case where $\mu>0 $.
 
It can be easily seen that for the proper boundary conditions, the solutions read (see \cite{Emrah} for details)
\be{2.29} 
\K=-1: \qquad \phi_{i} = -\frac{G_Nm_i}{\sinh\chi}\exp\left(-{\sqrt{\nu+1}\, \chi}\right), \quad 0<\chi<+\infty
\ee
and
\be{2.30} 
\K=+1: \qquad
\phi_{i} = -G_Nm_i \frac{\sinh\left[\sqrt{\nu-1}(\pi-\chi)\right]}{\sinh\left(\sqrt{\nu-1}\pi\right)\sin\chi},\quad 
0<\chi\leq \pi
\, .
\ee

\ 

The expressions for total gravitational potentials can be trivially found by adding the individual contributions:
\be{2.31}
\Phi = \sfrac{4\pi G_N}{c^4} \overline{\varepsilon}\lambda^2_{\rm eff} - \sfrac{G_{\!N}\,}{c^2 a}
\sum_i \sfrac{m_i}{\sinh\,l_i}\, \exp\left(-\sqrt{\nu+1}l_i\right)\, ,\quad \K=-1
\ee
and
\be{2.32}
\Phi = \sfrac{4\pi G_N}{c^4} \overline{\varepsilon}\lambda^2_{\rm eff} - \sfrac{G_{\!N}\,}{c^2 a}
\sum_i m_i \frac{\sinh\left[\sqrt{\nu-1}(\pi-l_i)\right]}{\sinh\left(\sqrt{\nu-1}\pi\right)\sin l_i}\, ,\quad \K=+1 
\ee
where $l_i$ represents the geodesic distance between the $i$-th mass and the observation point.


\section{Conclusion}

In the present paper we have investigated the effects of the peculiar velocities of inhomogeneities and the spatial curvature of the universe on the shape of the gravitational potential.  To this end, we have considered the spatially open and closed universe models, filled with cold dark matter in presence of the cosmological constant. The FLRW background metric is disturbed by inhomogeneities in the form of discrete point-like masses. Since we are interested in the associated gravitational potential, we have considered only scalar perturbations. We have reduced the system of linearized Einstein equations to an equation for the gravitational potential, in which we have effectively included the contribution of the peculiar velocities of gravitating masses. 
This equation is of Helmholtz type, where the Laplace operator is defined by the metric of the constant curvature space. 
It is well known that fields which satisfy the Helmholtz equation are characterized by the corresponding screening lengths. In our case, this is $\lambda_{\rm eff}$ defined by Eqs.~\rf{2.16}-\rf{2.18} .
It can be easily seen that $\lambda_{\rm eff}$ is determined by three separate contributions. The first one is due to the presence of the background matter with $\overline\rho \neq 0$ (see Eq. \rf{2.17}). The second one is defined by the peculiar velocities (see the first term on the right-hand side of Eq. \rf{2.18}). Finally, the third contribution is due to the spatial curvature of the universe  (see the second term on the right-hand side of Eq. \rf{2.18}).  

In a spatially flat universe, peculiar velocities significantly alter the value of the screening length \cite{EE}. For example, for the present time, the screening length is reduced from 3.74~Gpc to 2.57~Gpc when peculiar velocities are introduced in the calculations. Eqs. \rf{2.16}-\rf{2.18} show that a similar effect takes place also in a spatially curved universe. On the other hand, spatial curvature itself does not significantly affect the value of the screening length. For example, if we take $H_0=67.4\, {\rm{km}}\, {\rm{s}^{-1}}{\rm{Mpc}^{-1}}$ and 
$|\Omega_{\mathcal{K}}|=0.044$, then $3|\mathcal{K}|/a_0^2=3|\Omega_{\mathcal{K}}|H_0^2/c^2\approx 6.7\times 10^{-3}\,{\rm{Gpc}^{-2}}$. This value is considerably less than $1/(2.57)^2{\rm{Gpc}^{-2}}\approx 0.15\, {\rm{Gpc}^{-2}}$. However, even though introducing the curvature does not significantly alter the screening length, the sign of the spatial curvature does play a crucial role in determining the form of the gravitational potential (see Eqs. \rf{2.31} and \rf{2.32}). An additional important bonus of taking into account the peculiar velocities is that in a spatially closed universe, the gravitational potential is described by a single formula \rf{2.32} throughout the entire relevant period of the evolution. If we neglect the peculiar velocities, we need to consider three different expressions for the gravitational potential, related to three different stages of the evolution, and of course, smoothly connect these solutions  to one another \cite{Emrah}.


\section*{Acknowledgments}

The work of M. Eingorn and A. McLaughlin II was supported by the National Science Foundation HRD Award number 1954454.


\end{document}